\documentclass[11pt,a4paper,english]{article}
\usepackage{graphicx}
\usepackage[colorlinks=true, linkcolor=blue, citecolor=blue, urlcolor=blue]{hyperref}
\usepackage{comment}
\usepackage{setspace}
\usepackage{enumitem}
\usepackage{xspace}
\usepackage[dvipsnames]{xcolor}
\usepackage[most]{tcolorbox}
\usepackage{graphicx}
\usepackage{lipsum}
\usepackage[margin=1in]{geometry}
\usepackage{fancyhdr}
\usepackage{authblk}

\title{Challenges in Human-Agent Communication} 

\author[1]{Gagan Bansal\thanks{Equal contribution.}}
\author[1]{Jennifer Wortman Vaughan$^*$}
\author[1]{Saleema Amershi}
\author[1]{Eric Horvitz}
\author[1]{Adam Fourney}
\author[1]{Hussein Mozannar}
\author[1]{Victor Dibia}
\author[2]{Daniel S. Weld$^*$}
\affil[1]{\small{Microsoft}}
\affil[2]{\small{Allen Institute for Artificial Intelligence}}
\affil[ ]{\small{\texttt{\{gaganbansal, jenn, samershi, horvitz, adam.fourney, hmozannar, victordibia\}@microsoft.com}, \texttt{danw@allenai.org}}}

\date{}

\begin{document}

\maketitle

\newcommand{\bug}
    {\mbox{\rule{2mm}{2mm}}}
    
\newcommand{\DSW}[1]
    {\bug \footnote{\textcolor{ForestGreen}{\textit{DSW: #1}}}}
\newcommand\dan[1]{\textcolor{ForestGreen}{#1}}

\newcommand{\GB}[1]{
{\bug \footnote{\textcolor{BurntOrange}{\textit{GB: #1}}}}}

\newcommand{\jenn}[1]{
{\bug \footnote{\textcolor{magenta}{\textit{Jenn: #1}}}}}

\newcommand{\eric}[1]{
{\bug \footnote{\textcolor{red}{\textit{Eric: #1}}}}}

\newcommand{\adam}[1]{
{\bug \footnote{\textcolor{blue}{\textit{Adam: #1}}}}}

\newcommand{\saleema}[1]{
{\bug \footnote{\textcolor{purple}{\textit{Saleema: #1}}}}}

\newcommand{\etal}{et al.\xspace}

\begin{figure}[h]
    \centering
    \vspace{-2em}
    \includegraphics[width=\linewidth]{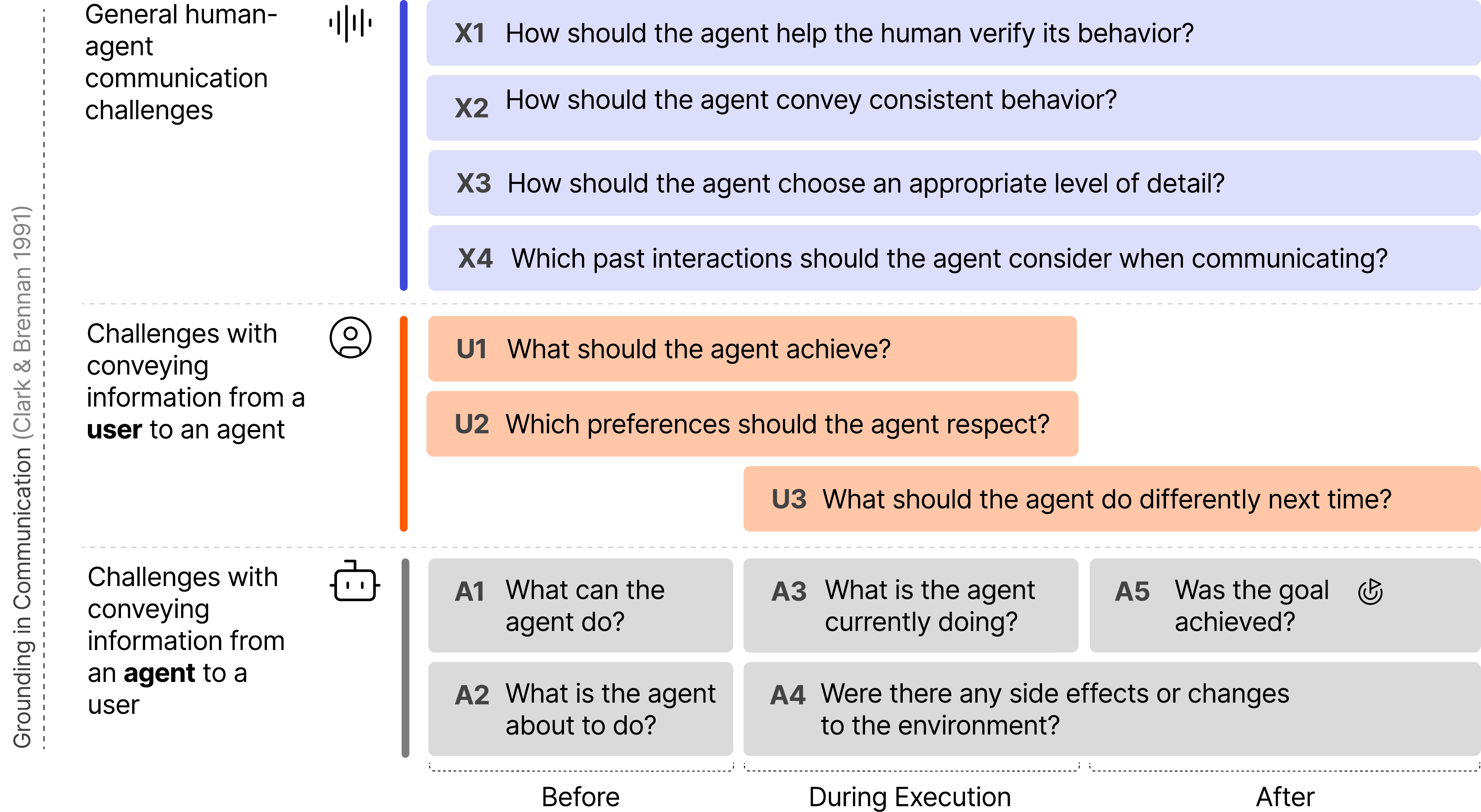}
    \caption{Modern agents introduce new (or newly complex) challenges for human-agent communication. We group these challenges broadly into three categories: those that pertain to information agents need to convey to users (A1--A5), those that pertain to information users need to convey to agents (U1--U3) and general difficulties with human-agent communication (X1--X4). These challenges may be differently applicable across the phases of human-agent interaction (x-axis): before, during, and after an interaction/execution.  }
    \label{fig:landing}
\end{figure}

\begin{abstract}
  Remarkable advancements in modern generative foundation models have enabled the development of sophisticated and highly capable autonomous agents that can observe their environment, invoke tools, and communicate with other agents to solve problems. Although such agents can communicate with users through natural language, their complexity and wide-ranging failure modes present novel challenges for human-AI interaction. Building on prior research and informed by a communication grounding perspective, we contribute to the study of \emph{human-agent communication} by identifying and analyzing twelve key communication challenges that these systems pose. These include challenges in conveying information from the agent to the user, challenges in enabling the user to convey information to the agent, and overarching challenges that need to be considered across all human-agent communication. We illustrate each challenge through concrete examples and identify open directions of research.  Our findings provide insights into critical gaps in human-agent communication research and serve as an urgent call for new design patterns, principles, and guidelines to support transparency and control in these systems.
\end{abstract}

\section{Introduction}

Artificially intelligent \emph{agents} are systems that can autonomously perceive and take actions in an environment~\cite{russell-norvig}. While the study of AI agents traces back many decades~\cite{russell-norvig,shoham-leyton-brown,Xi2023rise}, recent advances in generative foundation models that can output novel text or images based on natural language prompts have paved the way for the widespread development and deployment of a new class of agents that are increasingly sophisticated, powerful, and general purpose.  Granted the ability to access the internet,\footnote{\url{https://copilot.microsoft.com/}} connect with other applications through APIs,\footnote{\url{https://platform.openai.com/docs/guides/function-calling}} and even generate and execute computer code,\footnote{\url{https://platform.openai.com/docs/assistants/tools/code-interpreter}} today's AI agents can perform actions such as scheduling meetings, booking flights,\footnote{\url{https://www.expedia.com/newsroom/expedia-launched-chatgpt/}} ordering food,{\footnote{\url{https://aka.ms/magentic-one-blog}}} or purchasing groceries, taking actions that impact both the digital and physical realms. Ongoing developments in multi-agent architectures are further expanding the capabilities and use cases of agents~\cite{autogen,Xi2023rise, fourney2024magentic}. \looseness=-1

Along with agents' greater capacity to take actions in the open world and to complete goals on behalf of users comes a wider range of potential failure modes and associated costs \cite{ShavitPracticesGoverningAS,zhang2024interactionimpacts,gabriel-arxiv2024}. In fact modern applications are extending agent-centric activities into high-stakes scenarios. For example, an agent that can shop on a user's behalf can spend money in unintended ways or inadvertently leak the user's address, credit card number, or other sensitive information.  An agent that can execute computer code can corrupt files, alter important settings, overwrite family photos or work assignments, and take actions that jeopardize security.  Without a proper understanding of an agent's capabilities and limitations and the ability to verify its actions, a user may over-rely on an agent, leading to the user requesting that the agent perform a task that it is incapable of completing.  Particularly in situations where failures are costly or likely to occur, it is critical to build agents that allow users to clearly express their goals, preferences, and constraints to the agent and to form an accurate mental model of how the agent will behave.  Users should also be able to monitor the agent's behavior and effectively guide the agent with feedback and corrections as needed. Put another way, to enable effective collaboration with users, agents and systems of multiple agents must be designed to support \emph{transparency} and \emph{control}. \looseness=-1

The key to enabling transparency and control is effective two-way communication aimed at establishing \emph{common ground} about the user's goals (e.g., as represented by the \emph{content} of the user's request) and about the \emph{process} the agent intends to  take to achieve these goals~\cite{clark-1991,Satyanarayan2024Intelligence}. Achieving common ground is an activity that begins with the user's first introduction to the agent and continues throughout and after usage. Communication between users and agents can help a user to determine when and how much to rely on the agent and, through the iterative nature of dialog about intentions, capabilities, and activities, can help the user to identify and correct misconceptions before irrevocable actions have been taken.

However, communication that establishes common ground can be difficult.  Even with small-scale or special-purpose models, it can be difficult to characterize an AI system's abilities in a way that users understand~\cite{DK17,L18,vaughan2021humancentered,liao2021human, mozannar2024effective}. Generative foundation models have characteristics that make transparent disclosure of their operation particularly challenging, including the wide and constantly evolving range of tasks they can perform at different levels of competency, their massive and opaque architectures, sensitivity to prompt and context, the stochasticity of their output, and the diversity of their user bases who may require different levels of detail~\cite{liao2024transparency}. The challenge of understanding agents' operation is further exacerbated by the complexity of agentic workflows, in which tasks may be decomposed and carried out over multiple steps---sometimes by multiple interacting agents with different roles, privileges, and access to different information~\cite{autogen,fourney2024magentic}.  While the literature on human-AI interaction offers general guidance and sets of principles~\cite{horvitz-chi99,amershi-chi2019,pairguidebook}, interacting with systems of one or more autonomous, tool-using AI agents raises new transparency and communication challenges that have yet to be addressed. \looseness=-1

We present a set of challenges that arise in the process of establishing common ground between human users and AI agents, as summarized in Figure~\ref{fig:landing}. We arrived at these challenges based on our experiences building and experimenting with complex AI agents and multi-agent systems, drawing on the literature on human-AI interaction and collaboration, including prior work on establishing common ground between people and machines. Some of these challenges reflect the difficulty of conveying necessary information from the agent to the user to allow the user to form an appropriate mental model of the agent and monitor its behavior. Some reflect the difficulty of designing the agent to enable the user to convey their own goals, preferences, and constraints to the agent and guide the agent with feedback.  Others are overarching challenges reflecting general difficulties with communication, such as avoiding inconsistencies and reducing the burden on users.  While this list isn't exhaustive, we hope it serves as a starting point for discussion and future research.

We note that many parallels can be drawn between our proposed challenges and those faced in establishing communication and coordination within teams of humans (summarized in Section~\ref{sec:previous}).
We use the term ``common ground''  in the sense of Brennan \cite{Brennan2000TheGP} to refer to operational alignments, including on shared inferences such as the likelihoods of the current state of the world or future outcomes jointly considered by the human and agent, rather than broader cognitive or experiential similarities and do not suggest that agents understand in a similar way to humans.

\subsection{What is different about modern AI agents?}

While the agent-based perspective in AI is not new, two primary technical advances distinguish today's agentic applications from those of the past. First, unlike previous AI agents, which were typically based on simpler models with well-structured inputs and outputs, today's are based on \emph{generative} foundation models, like large language models or multimodal neural models, that can output novel text and/or graphical content based on natural-language (or multimodal) prompts or commands.
These generative foundation models have demonstrated remarkable performance across a wide range of use cases or tasks~\cite{bubeck-arxiv2023} and emerging capabilities continue to be discovered as models improve~\cite{wei2022emergent}. The broad abilities of foundation models enables the design of agents that exhibit a wide range of capabilities that often, though not always, include language understanding and generation competencies that allow people and agents to communicate through natural language.

Second, generative foundation models can be given the ability to invoke a wide range of resources or tools, for instance, through APIs~\cite{schick-neurips2024}.
Such tools and plugins enable agents to interact with the world, facilitating actions across domains such as including finance, communication, and physical activities in the open world.
Some agents can even execute arbitrary Python code in various environments, including cloud-based containers, local systems, or the user's own execution environment. These tools serve as sensors that enable the agents to perceive aspects of the world and effectors that allow them to take actions resulting in changes to world state. For instance, one tool might enable the agent to fetch weather data from a web API, while another might allow the agent to plot the data and save it to the user's local or cloud storage.

For another example, consider Devin, an ``AI software engineer'' agent with access to a command line environment, code editor, and web browser~\cite{devin}. In fact, its ability to control a Web browser allows it to browse information on the Web and even purchase products or services---actions with significant potential side effects \cite{zhang2024interactionimpacts}.
Since Devin has basic reasoning and planning capabilities, it can execute (often successfully!) complex engineering tasks, such as debugging code or even training and fine-tuning new AI models. However, current limitations in LLM planning~\cite{kambhampati2024can} and other issues such as hallucinations mean that Devin's actual behavior cannot be perfectly anticipated, potentially leading to costly or catastrophic errors.

In summary, generative language models empower tool-using capabilities that increase the power of agentic systems radically, allowing them to execute meaningful actions in the world and creating new (or newly complex) challenges as discussed in the rest of the paper.

\subsection{Outline}

We consider a set of emergent challenges pertaining to {\em human-agent communication}. While the rise of new degrees of autonomy and use of tools by agents poses many other important challenges including technical~\cite{Ji2022SurveyOH,Kambhampati2020ChallengesOH,Ji2023AIAA,Valmeekam2023OnTP,compound-ai-blog,dibia2024autogen}, ethical~\cite{chan2023facct,lazartutorial,gabriel-arxiv2024}, safety~\cite{Amodei2016ConcretePI,ShavitPracticesGoverningAS}, and fairness~\cite{Jobin2019TheGL,Madaio2020CoDesigningCT} challenges, we consider these to be outside the scope of this paper.

We start our discussion with a recap of previous work from the perspectives of psychology, cognitive science, AI, and HCI. In Section~\ref{sec:general}, we list four overarching challenges that apply across all human-agent communication. In Sections~\ref{sec:usertoagent} and \ref{sec:agenttouser}, we discuss issues that emerge from the need to communicate information from the agent to the user and user to the agent, respectively.   In each section, we describe the challenges in detail, contextualize them with respect to the literature, and provide concrete examples from existing systems and emerging applications. We outline possible steps towards solutions for each challenge and identify open research questions. Finally, in
Section~\ref{sec:conclusion}, we close with a call to action for research in human-agent communication.

\section{Previous Work}
\label{sec:previous}

We now summarize some of the relevant literature that we build upon, drawing from prior research and results in psychology, cognitive science, AI, and HCI. 

Researchers in philosophy and psychology have long studied human communication and teamwork.  The notion of grounding in communication is a central concept proposed by Clark and Brennan that embodies the collection of {\em mutual knowledge}, {\em mutual beliefs}, and {\em mutual assumptions} that are essential for communication between two people~\cite{clark-1991}.  To effectively ground their communication, the participants must ``coordinate both the content and process'' of each conversation. The process of achieving a shared mutual belief (common ground) is called {\em grounding}. Clark proposed successful grounding as requiring the parallel action of multiple levels of analysis, including the establishment of a communication \emph{channel}, the exchange of recognized \emph{signals} across the maintained channel, the interpretation of \emph{intentions} via decoding meaning in the signals, and the control of the back and forth of an effective \emph{conversation} with contributions and clarifications being made by dialog participants.

Drawing parallels to human communication, Brennan applied grounding theory to human-computer interaction, where, as in human communication, ``people need to be able to seek evidence that they have been understood and to provide evidence about their own intentions'' \cite{Brennan2000TheGP}.  In this case, the term ``common ground'' refers to operational alignments rather than cognitive or experiential similarities.  Brennan noted that poor feedback or affordance mechanisms often result in users needing to execute laborious checking actions to achieve common ground. The breadth of capabilities (and stochasticity) displayed by generative language models exacerbate these problems, as we discuss.  

In related work informed by research in psychology on human-human communication, explicit computational grounding machinery was developed and integrated within computational architectures to support human-AI grounding processes as \emph{joint activity} aimed at achieving mutual understanding \cite{hp1999}. Another effort explored the development of computational analogs of the multiple levels of coordination for grounding proposed by Clark, including inferences at channel, signal, intention, and conversation levels, to establish common ground between a user and a dialog system that could execute actions in the open world. \cite{paek2000}. In later work, definitions of grounding in human-AI collaborations were shifted to an expected utility framework. In the approach, grounding on ideal actions for an agent is guided by uncertainties that an agent has about a user's goals and the costs and benefits of different actions \cite{paek2000}. The latter work also explored the use of visual signaling by the agent to communicate to the user the degree to which an agent believed it was grounded with the user via display of smoothly changing colors. More recent studies have explored grounding between machine and human in a multimodal setting, where uncertainties that an agent has about a user's goals and intentions are inferred via signals drawn from multiple streams of information, including language, vision, predictions about states of the world, and conversational flow, and then communicated to users via dialog acts and gestures \cite{Pejsa2014}.   

Several efforts have focused on establishing common ground between machines and people on memories about activities that have occurred in the past. Notions of \emph{shared memory} have been constructed via mechanisms that infer important milestone events drawn from larger streams of events observed over time that are likely to be memories accessed by, referred to, or assumed in joint activities by users when interacting with AI systems. Studies have explored the identification of memory milestones to enable easy reference to events by users and AI systems \cite{ringel2003milestones,SharedMemory2004} and shared, referable memories of prior interactions and situations of the type that people would have with others whom they have worked with over time \cite{rosenthal2013execution}.   

Among people, grounding typically proceeds in alternating phases of {\em presentation} and {\em acceptance} with gestures and facial expressions often serving as acknowledgments~\cite{Goodwin1981ConversationalOI}. Communication made across other media (e.g., when communicating with or via computers) forces people to use more elaborate grounding methods and often increases communication costs for both the speaker and addressee~\cite{clark-1991}. These media-based constraints play into the communication challenges discussed in this paper.

Dennis et al. elaborated on the theory of communication processes and distinguish between {\em conveyance} and {\em convergence} processes. They argued that the former can do with lower {\em media synchronicity} whereas convergence benefits from higher media synchronicity~\cite{Dennis2008MediaTA}. Here, media synchronicity indicates the degree to which the communication medium allows coordinated synchronous behavior. They conclude that communication is improved when participants use various media to perform a task. Generalizing `media' to computational affordances, these theories also inform our proposed solutions.

The study of human-agent collaboration should also be informed by social and cognitive psychologists' work on decision-making processes of groups of people, which may often surpass the performance of  individuals comprising the group. Many of these prior studies corroborate a theory that groups collectively perform better than the average (and many times, best) individual on many types of problem-solving tasks~\cite{hill_1982_group,carey_2012_groups}. These effects were more strongly seen on some tasks than others and attributed to the {\em demonstrability} of proposed solutions to the task~\cite{laughlin_1986_demonstrability}---which may be seen as properties of the group and task that facilitate grounding. Some of our suggestions for X1--X3 benefit from these insights from psychology and their adaptations into principles for AI-generated explanations~\cite{Fok2024InSO}.  Shared mental models, closed-loop communication, and mutual trust have been identified as core mechanisms in successful human teamwork~\cite{Burke2005IsTA}, and some of these aspects map clearly to human-AI teams~\cite{Bansal2019BeyondAT}. Still, the differences in understanding, decision-making processes, and communication styles between humans and machines change the context and introduce new challenges.

Our work also builds on significant recent work on design guidelines for human-AI interaction~\cite{amershi-chi2019,pairguidebook,Shneiderman2020HumanCenteredAI,Yang2020ReexaminingWW,Xu2021TransitioningTH,Kambhampati2020ChallengesOH,Yildirim2023InvestigatingHP}. The rise of generative AI models has led to a massive shift in the ways AI is being deployed---from interactive classifiers into powerful tool-using agents. This transformation raises a number of new issues. For example, using tools allows agents to take a much wider range of actions~\cite{Schick2023ToolformerLM}, increasing complexity and the number of ambiguities that must be grounded. In contrast to previous AI applications, agentic systems commonly have an emphasis on generalist capabilities~\cite{Reed2022AGA,Wang2024OpenHandsAO}, which likewise raises the complexity of grounding. These systems are often crafted as ``society of mind'' \cite{minsky1988society} agents that pass tasks to each other~\cite{autogen}, which can result in complex emergent behaviors---again posing challenges for grounding. As a result of these jumps in complexity, it is appropriate to revisit both guidelines and challenges for human-AI interaction. 

A closely related work is the report from Shavit et al \cite{ShavitPracticesGoverningAS} which offers a set of practices towards making the operation of agentic AI systems more safe and accountable. That work intersects with a subset of the challenges that we discuss in this paper; specifically challenge X1 and A3 are closely tied to ``Legibility of Agent Activity" and challenge X3 and A3 are tied to ``Constraining the Action-Space and Requiring Approval.'' Our work makes progress by also reflecting on broader challenges in human-agent communication, including how to effectively convey capabilities and limitations, manage information flow between agents and users, and address emergent issues in multi-agent systems.
\section{Overarching Challenges for Human-Agent Communication}
\label{sec:general}

Before digging into what specific information must be communicated between a human user and AI agent to establish common ground, we begin with a discussion of four overarching challenges that should be considered across all human-agent communication (challenges X1--X4 in Figure~\ref{fig:landing}).  We briefly introduce each here and return to them when discussing the challenges in communicating specific information in Sections~\ref{sec:usertoagent} and~\ref{sec:agenttouser}.

\subsection*{X1: How should the agent help the user verify its behavior?}

Modern agents are powerful but remain imperfect. As a result, when agents get tasked with goals, it is likely that they will make mistakes---especially when these goals are complex or multi-step. Consider an agent tasked with fixing a programming bug in a GitHub repository. This agent can fail in numerous ways. For example, it might fail to understand the issue and its requirements, such as the fact that any solution must be backward compatible; it might develop a plan that implements the fix in the source directory but neglects to update the test cases; it might propose a fix that is ``unsafe'' because it is vulnerable to exploits or propose a fix that is convoluted and time-consuming to review; it might propose a change that has side effects such as removing unrelated functionality.

To establish establish common ground, it is necessary not only to convey pieces of information but to \emph{verify} that a common understanding has been reached.  In the context of human-AI communication, it is not enough, for instance, for a user to state their goals and preferences if these goals and preferences are misunderstood or ignored by the agent.  It is, therefore, critical to enable the user to verify their understanding of the agent's plans and actions as well as the agent's understanding of their own goals, preferences, requirements, and feedback. Establishing mechanisms for the user to easily verify the agent's behavior is a necessary condition of effective communication~\cite{Fok2024InSO} and an important aspect of effective teamwork~\cite{Laughlin1980SocialCP}. Without this ability, the user may not be able to steer agent behavior to avoid costly mistakes or accomplish the goal. Furthermore, they would fail to develop an accurate mental model of the agent's capabilities or provide feedback to improve the agent's future performance.

\subsection*{X2: How should the agent convey consistent behavior?} 

A second overarching challenge for human-AI communication is how to prevent the agent from confusing the human with behavior or outputs that are (or at least appear) \emph{inconsistent}.
Inconsistency can arise for multiple reasons. One is the agent's behavior's inherent \emph{stochasticity}, which stems from two main sources: the probabilistic nature of the underlying foundation model's outputs, and the complex interaction patterns that emerge during task execution. Even with deterministic foundation model outputs (e.g., a model with temperature set to 0), when agents operate in dynamic environments, their orchestration logic---which governs action selection, inter-agent delegation, and task completion criteria---can produce different sequences of actions across runs as the environment changes in response to agent actions. While stochastic behavior may be desirable in some cases---for instance, if the user would like the agent to generate a variety of different options to choose from---in other cases, it can hinder the user's ability to create an accurate mental model of the agent. For example, consider an LLM-based agent that can create visualizations based on a high-level description from the user. When invoked with the same input multiple times, the agent might choose different libraries or tools to plot the visualization. Sometimes, it might write code to fetch the data from a hosted service; other times, it might generate that data from its memory.
It could also unpredictably apply different visual styles, which would be especially problematic if the user is unaware.

Perceived inconsistencies can arise for other reasons, for instance, when the user's mental model of the world does not match up with the information the agent is acting on. Even if an agent is taking actions that align with the user's goals, its actions may appear misaligned if the human's model of the world is different~\cite{Sreedharan2024PlanningWM}; think of a shopping agent purchasing what appears to be an overly expensive widget because it knows that the cheaper model is incompatible with the user's needs, but fails to consider the user's budget limitations.

If such forms of (perceived) inconsistencies are not minimized, users may become confused about whether the agent has accomplished their goal, whether it is capable of accomplishing the same (or a similar) goal again, whether it would follow similar approaches as it used in the past, and how to consistently and accurately steer its behavior. Over time, this will likely lead to diminished trust.

\subsection*{X3: How should the agent choose an appropriate level of detail?}

While it is essential to design agents in a way that enables users to verify their behavior and avoids confusing users, these needs must be balanced with the need to avoid burdening or overwhelming the user by providing \emph{too much} information.
Interactions with modern agents necessitate richer bidirectional communication than was necessary with traditional AI systems. This increased emphasis on communication arises because of the complexity of the underlying AI systems and the complexity of the tasks the agents are being asked to perform. While more detailed communication can help establish a shared understanding, it can become counterproductive when instructing the agent becomes burdensome or reviewing agent outputs becomes overly cognitively taxing. It also may overconstrain the agent, preventing it from exploring alternative solutions or recovering from failures.

When designing for human-AI communication, there are several questions: Can the user easily instruct the agent regarding their goals, constraints, and feedback, or does this require a lengthy and potentially unreliable process? (See also Vasconcelos et al.~\cite{vasconcelos-cscw2022}.)
Are the explanations of the agent’s behavior clear and understandable, or are they too complex to be useful? Additionally, does the agent need to confirm every request, or can it act with more autonomy in familiar situations?

Consider how an LLM-driven agent might assist a user with drafting emails. Suppose a user frequently asks the agent to create follow-up emails with a similar structure and tone for recurring scenarios, such as responding to meeting requests or sending reminders. An overly verbose agent might repeatedly ask for clarification on tone, length, or recipients for every request, even when the user has previously provided clear preferences for these aspects. Instead, an ideal agent would leverage its understanding of past interactions to streamline the process, only requesting additional details for novel or ambiguous scenarios. This avoids unnecessary repetition and enhances user satisfaction.

While this challenge isn't new \cite{Poursabzi2021manipulating}, the generative power of today's language models makes it omnipresent (as the example above illustrates). 

\subsection*{X4: Which past interactions should the agent consider when communicating?}

Context plays a key role in many aspects of human communication and computing (e.g., in search and recommendation) \cite{Halliday1989LanguageCA,Bellotti2001IntelligibilityAA}. Interactions with modern agents can especially involve rich and intricate contexts that both the user and agent may draw from to facilitate communication and understanding, e.g., past interactions containing lengthy generations and exchanges with the user, background information about the user, or observations from the environment and tool use.
For example, consider an academic research agent who assists a scholar. This agent could benefit from reading all of the scholar's papers so that it can make use of references to past results and experimental designs, storing and reusing prior discussions with the scholar about an ongoing project, or considering observations from experiments that it helped conduct. 
This context could grow unbounded as the agent works with a user over days, months, and eventually years.  Agents should use this past context to satisfy each user request, but as the context grows huge, how can we ensure the agents focus on what is relevant for interpreting the current command? (See also Liu et al.~\cite{liu-tacl2023}.) This may depend on the task the agent is performing or the specific question the user asks at a given moment.

A similar challenge occurs when deciding which information to retain from the agent's percepts, such as when a tool retrieves a large amount of data (e.g., 1000 PDF documents). These percepts, like the interaction context, can also generate large and complex contexts, further complicating the challenge of maintaining relevance and focus in its responses.

In some cases, the user may want to limit the context that the agent is allowed to rely on; for instance, if allowing the agent to rely on particular sources of information would compromise privacy. The user should always be able to restrict the use of background information when they prefer. Fortunately, this need has surfaced in many analogous situations, suggesting applicable design patterns. For example, most internet browsers support incognito mode, where cookies and other personally identifiable information are suppressed.  Similarly, browsers allow users to view, edit, and purge the history of past browsing. Social networks support users who switch between different personas.  We suspect that similar tools will be important for supporting human-agent interaction. Additionally, recent work on agent workflow memory \cite{wang2024agent} shows how agents can learn reusable task workflows from past experiences to improve their performance on complex, long-horizon tasks. This approach has demonstrated significant improvements in web navigation tasks across diverse domains, suggesting that explicit memory mechanisms could be particularly valuable for agents executing repetitive tasks.

\section{Information Flow from the User to the Agent}

In this section, we discuss three challenges (U1--U3) related to designing agents to enable users to communicate necessary information. The first two concern the user's desires: what users want the agent to achieve and preferences for how the agent achieves it. The third challenge pertains to feedback and helping the agent improve over time. While much of the information content covered by U3 overlaps with U1 and U2, the time and context differ, affecting communication.
For each issue, we discuss existing research, describe how these difficulties have changed for today's agents, provide concrete examples, and outline possible solutions.

\label{sec:usertoagent}
\subsection*{U1: What should the agent achieve?}

When agents assist with or carry out tasks for people, one of the most critical pieces of information a user needs to communicate to the agent is the \emph{goal}---what they would like the agent to achieve. If the agent misunderstands the goal, the consequences can be dire. An agent may waste compute, money, and time accomplishing the wrong (or worse, harmful or unsafe) thing, create side effects, or cause user frustration and abandonment. It is then critical that we ask how to design agents so that users can clearly express their needs and intent and resolve possible ambiguities~\cite{Dey2005DesigningMF}.

The challenges that arise in interpreting user intent are well studied in domains such as classical planning, personal assistants, mixed-initiative systems~\cite{horvitz-chi99}, and context-aware computing~\cite{Dey2005DesigningMF,Mankoff2000InteractionTF,Li2020MultiModalRO}.  For example, in classical planning, users may specify their goal via a set of logical conditions that are true in the goal state. Many formal languages for specifying the goal and scenario exist, such as the Planning Domain Definition Language and its variants~\cite{McDermott1998PDDLthePD,Younes2004PPDDL1,sanner2010relational}. These formal languages provide a precise and explicit way to define goals. While precise, these approaches are less intuitive and natural for users compared to specifying goals in natural language. One reason is that these formal languages inherit the limitations of first-order logic, such as its limited expressivity compared to natural language specifications, which is now common with modern agents.  In contrast, while more intuitive, goals specified in natural language introduce challenges such as ambiguity and imprecision.

While goals were specified in natural language in the personal assistant literature (e.g., {\em ``Set an alarm clock for 8 am tomorrow''}), there were many key differences: the specified goals were simple and limited to a small number of actions, an intent recognition system was used to map the intent to a ``skill'' (similar to function calling), and these systems did not adapt based on new observations e.g., based on results of a previous skill~\cite{budiu-Alexa-usability}.
With modern agents, people may express intents at a much higher level, and those intents may necessitate solutions that involve multiple steps of reasoning and tool use.
Natural language and dialog can often be incomplete, convoluted, or ambiguous, making correct interpretation more challenging. These differences can increase the chance that modern agents might misunderstand what the users want the agents to achieve.

One important mechanism that could improve communication of goals from the user to the agent is detecting and resolving critical points of uncertainty in the agent's understanding of the goal~\cite{horvitz-chi99}. 
For example, if the user asks a research assistant agent to see top papers by ``Peter Clark,'' there may be many potential authors with that name. Disambiguating this request may be seen as a process of finding common ground. 
This disambiguation might be made explicit in some cases, but risks overburdening the user (X3)~\cite{Keyvan-10.1145/3534965}. Alternatively, the agent could disambiguate between alternate goals using available context (X4)---again, a type of common ground. For example, rather than asking which Peter Clark, the agent could choose the one in the same discipline as the user or the one the user has cited in the past.

\begin{tcolorbox}[breakable, colback=orange!5!white, colframe=orange!75!black, title=U1. What should the agent achieve?]

\paragraph{Example 1: Travel Itinerary Planning} 

A user instructs an agent to plan a business trip, specifying the destination, preferred airlines, and meeting times. However, the agent misinterprets the trip's purpose, planning it as a leisure vacation instead. It books scenic flights with multiple layovers, selects a luxury resort far from the meeting venue, and even schedules irrelevant activities like guided tours. As a result, the user ends up with an itinerary that fails to meet their professional needs, potentially leading to missed meetings and wasted time. This highlights the importance of clear goal communication and the need for iterative verification and refinement to ensure the agent’s plans align with the user's intent.
    
\paragraph{Example 2: Scientific Literature Search} 

We asked Microsoft CoPilot\footnote{\url{https://copilot.microsoft.com}} and ChatGPT\footnote{\url{https://chatgpt.com}} to find recent papers that cite a previous paper~\cite{autogen}. Unfortunately, both systems misunderstood the goal and returned an unsatisfactory answer. CoPilot returned the links to the previous paper itself, whereas ChatGPT returned only two papers, even though the paper in context has been cited by a few hundred papers.
    To fix this, a natural next solution for the end-user may be {\em goal refinement} through dialog, e.g., specifying additional details like whether self-citations count and the minimum number of papers to fetch.
    \begin{center}
        \includegraphics[width=0.49\textwidth]{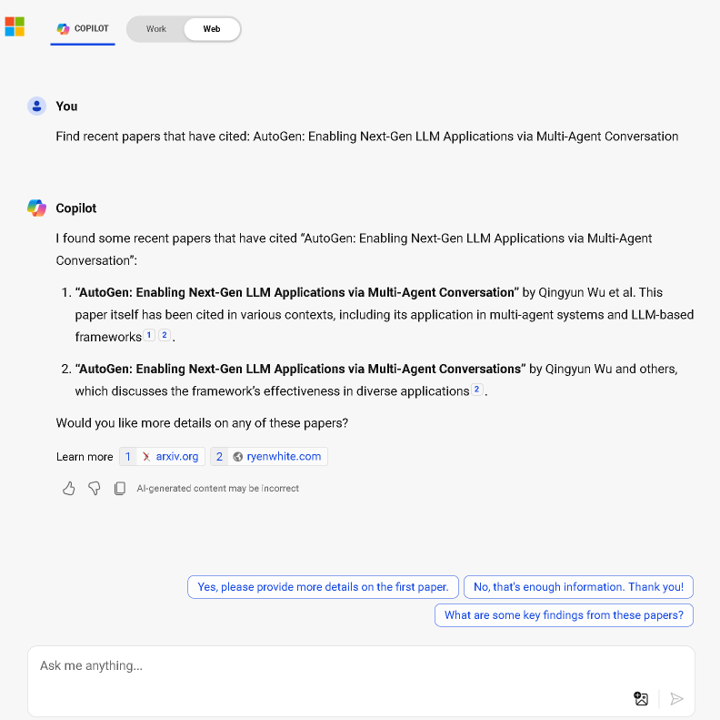}
        \includegraphics[width=0.49\textwidth]{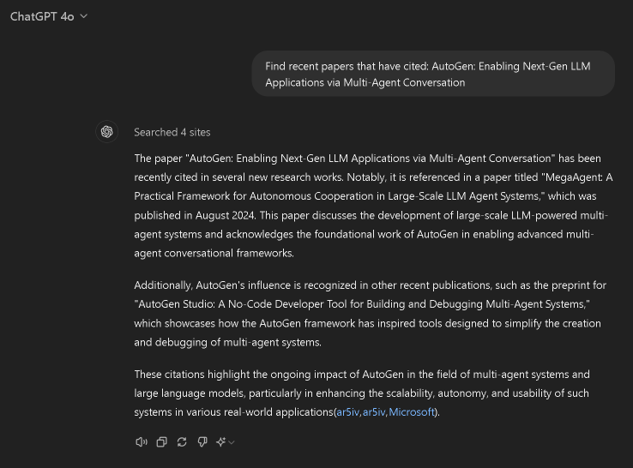}
    \end{center}

\end{tcolorbox}
\subsection*{U2: What preferences should the agent respect?}

For a given high-level, complex goal from the user, many possible ways to achieve the goal typically exist, but some plans are better than others.  For example, the user likely prefers to minimize time, expense, and harmful side effects. However, different users will have different notions of harm and, hence, different preferences.  Thus, understanding the user's preferences is a key component of common ground. 
How can the agent help the user clearly express their preferences, especially when they differ from standard norms? 

Steering agents to respect user preferences is related to efforts to align or fine-tune large language models toward specific behaviors, tasks, or domains. For example, techniques such as Direct Preference Optimization (DPO) were developed to steer LLM generations toward those that better align with user preferences and desired outcomes~\cite{Rafailov2023DirectPO}. However, while effective, DPO primarily focuses on aligning models to aggregate-level preferences, which may overlook nuances in individual user needs or fail to generalize well to diverse or context-specific preferences.

For generative agents that can carry out complex actions in the world, ensuring they respect user preferences and constraints can be even more challenging. For example, suppose the user wants to constrain the agent to avoid harmful behavior. In open-world settings where generative agents can operate, it may be impossible to precisely define ``harm'' or list all possible harmful behaviors~\cite{etzioni-aaai1994}. Further, generative models can discover novel solutions that contain previously unthought (or unspecified) harms~\cite{amodei-arxiv2016}. Tool use further increases the scope of possible harm by allowing agents to impact the environment, making the communication challenge even more salient. \looseness=-1

One possible solution to this challenge is to make inferred preferences and constraints visible or accessible to users. Some existing works use this approach; for example, the current version of ChatGPT displays a set of facts 
about the user it has memorized. It also allows users to modify (e.g., delete) these facts.\footnote{\url{https://openai.com/index/memory-and-new-controls-for-chatgpt/}} These methods are good steps towards creating common ground.

\begin{tcolorbox}[breakable, colback=orange!5!white, colframe=orange!75!black, title=U2. What preferences should the agent respect?]
    
    \paragraph{Example 1: Privacy Concerns in Data Sharing}

A user instructs an agent to handle a project involving sensitive personal data, explicitly specifying that this data should not be shared or stored in any external systems. However, the agent misunderstands or fails to properly recognize the importance of this privacy constraint. As a result, the agent inadvertently includes the sensitive data in a shared document or public database, exposing it to unauthorized parties. This data breach could lead to privacy violations, legal repercussions, a significant loss of trust from the user, and potential harm to individuals whose data was exposed. This scenario underscores the critical need for agents to understand clearly and strictly adhere to user-defined privacy constraints, ensuring that sensitive information is always protected. Additionally, it highlights the importance of iterative verification and refinement, where the user may need to review and correct the agent’s actions to prevent future breaches, emphasizing the challenge of accurate constraint communication.

\paragraph{Example 2: Web Scraping with Ethical Considerations} A user directs an agent to collect data from a specific website with clear instructions to avoid scraping sensitive information or violating the site’s terms of service. However, the agent misinterprets or overlooks these ethical constraints and proceeds to scrape restricted data, sensitive data, or data explicitly forbidden by the website’s terms. As a result, the agent gathers information the user never intended to collect, potentially leading to legal repercussions, ethical breaches, and damage to the user’s reputation. This example underscores the critical need for agents to understand and adhere to ethical guidelines and user-defined constraints when performing web-based tasks. It also highlights the importance of ensuring that agents recognize and respect boundaries, particularly when data privacy and legal compliance are at stake.
\end{tcolorbox}
\subsection*{U3: What should the agent do differently next time?}

Even if an agent develops an excellent  understanding of the user's goal, preferences, and constraints, in practice, it is possible that the agent will still make mistakes. For example, it may continue to use sub-optimal plans or incorrect tools because its planning is imperfect. Further, it is also possible that the agent continues to make wrong assumptions about the goals and preferences---perhaps because the user's tastes evolve. Agents will likely learn autonomously through mechanisms that enable them to invoke new tools, update memory, or finetune the underlying LLM~\cite{Schick2023ToolformerLM}, but we focus on scenarios where the user provides feedback.

This challenge is related to the problem of learning in recommender and interactive machine learning systems. For example, music and movie recommender systems may use implicit or explicit signals, such as dwell times or ratings, to improve future recommendations. Similarly, many end-user interactive machine learning systems focused on lightweight feedback mechanisms to help users steer classifier training and improve future predictions. 
Recent work on interactive prompt editing and refinement has also focused on enabling end users to improve prompts for LLM-based systems.
There is also work in the classic agent literature which has focused on improving feedback from users to agents. For example, Russell and Grosof focused on learning (first-order logic-based) concepts from examples that can be accommodated into the agent's learning process~\cite{russell-aaai1987}.

In contrast to traditional machine learning systems, generative agents can revise their behavior based on natural language text. While this allows for rich feedback, it also introduces the challenge of appropriately interpreting free-text feedback. How can we help users effectively express feedback to steer agent behavior?

\begin{tcolorbox}[breakable, colback=orange!5!white, colframe=orange!75!black, title=U3. What should the agent do differently next time?]
    
    \paragraph{Example 1: Code Debugging} A user asks an agent to write a Python function that sorts a list of customer orders by date and then filters out orders from non-preferred customers. The agent generates the code, but a bug causes the function to incorrectly filter out orders from some preferred customers when their names contain special characters. The user identifies the issue and provides feedback, specifying that the function should correctly handle names with special characters. The agent attempts to fix the bug but only partially succeeds---it correctly processes names with special characters but now fails to sort the orders by date in descending order. The agent struggles to integrate both fixes despite further feedback, leading to repeated debugging cycles. This scenario highlights the challenges in guiding an agent to learn effectively from feedback in complex coding tasks, where multiple aspects of the task need to be addressed simultaneously.

    \paragraph{Example: Travel Itinerary Adjustments}
    A user asks an agent to plan a week-long vacation, including flights, accommodations, and activities. The agent creates an itinerary that includes late-night events before early-morning tours and overlooks prioritizing the user's preference for cultural experiences over recreational ones. After completing the trip, the user provides detailed feedback---emphasizing the importance of scheduling adequate rest between activities, highlighting a preference for cultural activities, and requesting that the agent avoid certain types of accommodations in future plans. The agent must be able to use this feedback to refine its understanding of the user’s preferences for subsequent trip planning, showing improvement in aligning with these preferences over time.

\end{tcolorbox}

\section{Information Flow from the Agent to the User}
\label{sec:agenttouser}

In this section, we discuss five challenges related to the information flow from the agent to the user. These challenges focus on communicating information about the agent's capabilities, the agent's current and planned future actions, whether the user's goals that have been achieved, and any side effects that have occurred. We provide examples to illustrate each challenge and discuss potential solutions.

\subsection*{A1: What can the agent do?}

If a user does not fully understand the capabilities or limitations of an agent, they will not be able to make informed decisions about when and how to best use its assistance or what to expect when they do. Work with earlier AI systems has shown why addressing this challenge is crucial. As one example, Cai \ et \cite{cai-cscw2019} found that clinicians wanted clear information upfront about the basic features of AI models. They wanted to know what the models were good at, where they struggled, what perspectives they might have, and what they were designed to achieve. In the absence of such transparency about AI systems, people sometimes resort to informal experimentation to try and understand their capabilities and limitations~\cite{lugar2016BadPA}, which may lead to incomplete or inaccurate mental models ~\cite{eslami2016FolkTheories,Bansal2019BeyondAT}.

Several approaches have been proposed to help convey the capabilities and limitations of AI systems to people.  
Mitchell \ et al. \cite{mitchell-fact2019} introduced the idea of model cards, short documents that accompany trained machine learning models. Among other things, a model card may include details about how the model was trained, intended or unintended uses, evaluation results (potentially broken down by demographic), and ethical considerations. Companies including Google, Hugging Face, and OpenAI have adopted model cards, and different, more interactive formats for model cards have been proposed~\cite{Crisan2022interactive}.
However, a model card may be too long or technical to be practically useful for casual users, especially one intended to cover the range of behavior of a modern generative AI model.  (For example, the model card for GPT-4 is over 60 pages long!\footnote{\url{https://cdn.openai.com/papers/gpt-4-system-card.pdf}})  For end users, considering lighter-weight solutions to introduce capabilities and limitations that are integrated into an agent's interface or behavior may be more appropriate than providing documentation of the agent or model it is built on, which the user must look up outside of interactions.

Making clear what a modern agent can and cannot do requires answering many questions beyond what we see for traditional classification or recognition systems, such as: Which information does the agent have access to? How will the agent use this information?
Can the agent make permanent changes to the environment? Can the agent be interrupted without side effects? Or consider 
an agent that can write and run code. What languages and frameworks does the agent specialize in? Does it have access to the Internet? Can it use libraries that do not appear in the data used to train its underlying LLM, for instance, through RAG? Can it execute code that requires elevated privileges?  

We note that since an agent's capabilities and limitations may be updated over time, conveying these to users is not something that can happen only once.  It may become necessary to update the user on changes to what the agent can do over time \cite{amershi-chi2019}. 

\begin{tcolorbox}[breakable,colback=gray!5!white, colframe=gray!75!black, title=A1. What can the agent do?]

\paragraph{Example 1: Dataset Visualization}

The figure below shows the output of ChatGPT when asked to plot a chart of features from the UCI leaf dataset. The system uses GPT-4 and its code interpreter tools to generate the visualization. However, the visualization is based on simulated data, as shown on the left. This unexpected behavior raises a question: Why did the agent not use its known capability to browse the web to retrieve the actual dataset? Further investigation reveals that the agent used simulated data because it cannot access the Internet through the command line.

\begin{center}
    \begin{minipage}{0.49\linewidth}
        \centering
        \includegraphics[width=\linewidth]{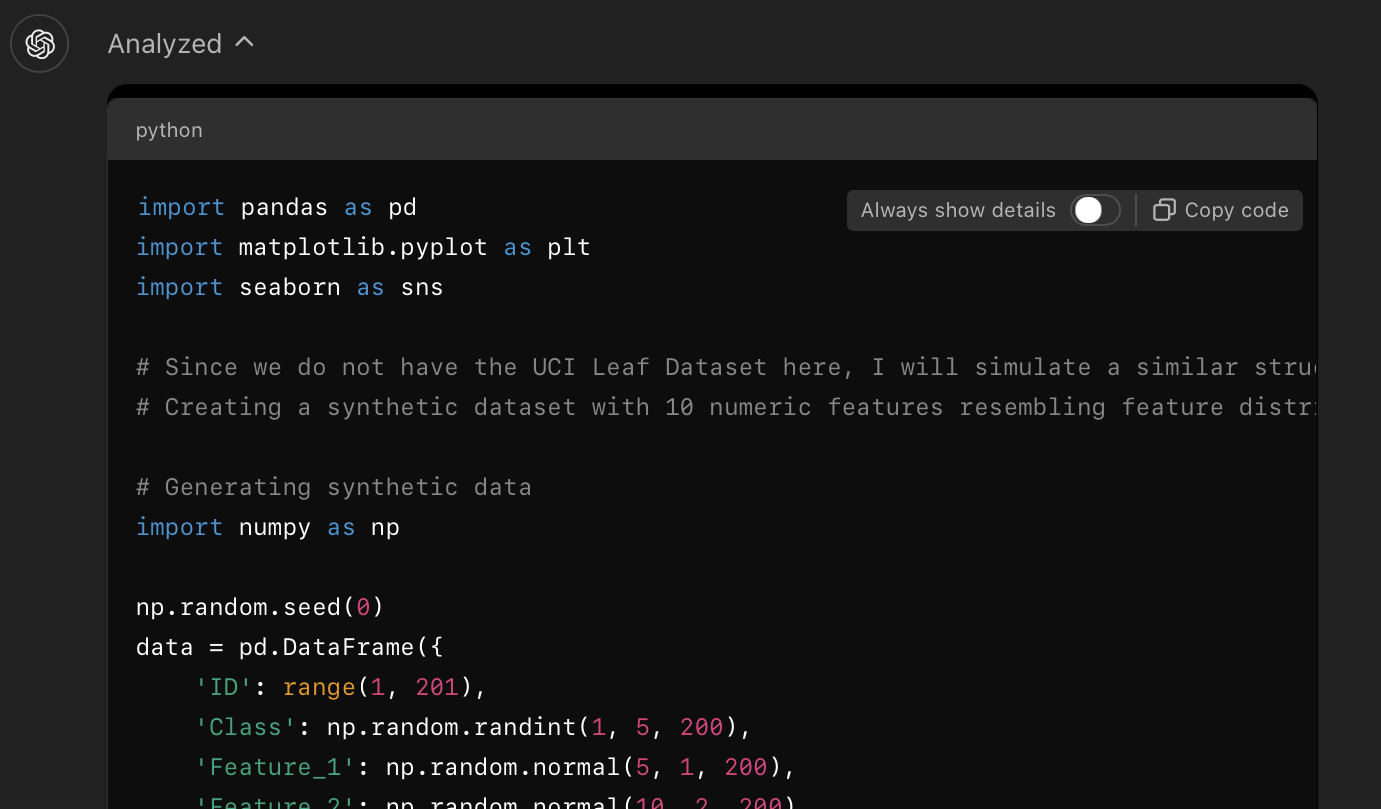}
    \end{minipage}
    \hfill
    \begin{minipage}{0.49\linewidth}
        \centering
        \includegraphics[width=\linewidth]{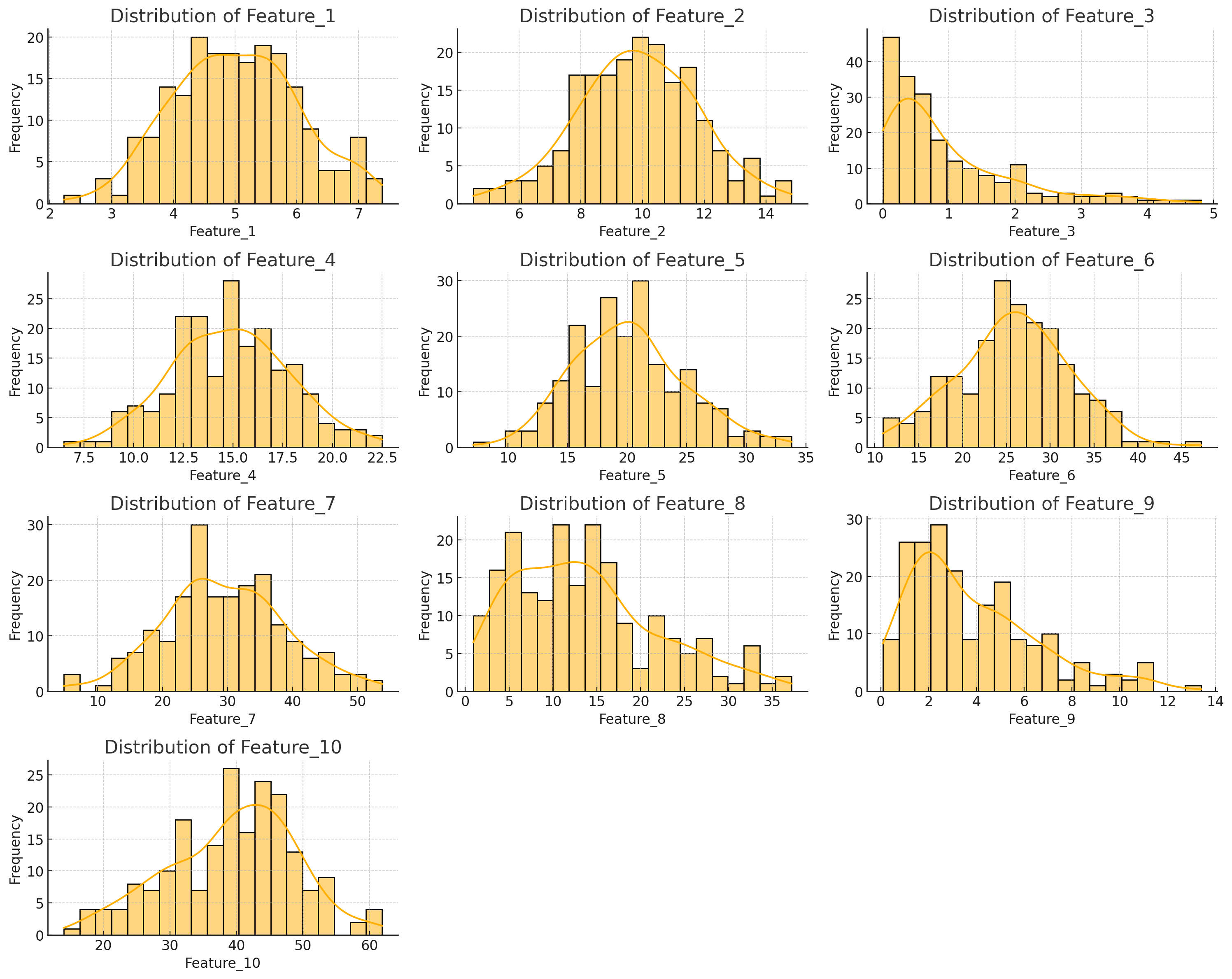}
    \end{minipage}
\end{center}

In this case, understanding that the agent can sometimes browse the web is insufficient. The user needs to understand the conditions under which this is possible.  Such information could potentially be laid out in an ``agent card'' (akin to a model card), but enumeration of all possible capabilities in appropriate detail may be impractical. A lighter-weight solution could indicate what tools or data the agent can access, but this may not be enough for the user to understand the nuances.


\paragraph{Example 2: Proving Theorems}
Consider an agent designed to assist with proving mathematical theorems. Suppose the agent is asked to prove the same theorem about prime numbers on two different occasions. Despite using the same underlying model and tools, the agent produces different proofs each time. The first time, it generates the proof directly without verifying its correctness. The second time, it generates a proof and then verifies it using a theorem prover. This inconsistency can confuse users about what the agent's capabilities are.  In this case, can we say that the agent can reliably prove theorems? The question is complicated by the agent's stochastic behavior and the various factors influencing it.

\end{tcolorbox}

\subsection*{A2: What is the agent about to do?}

To achieve a given complex goal, an agent may execute a large number of actions step-by-step. Before executing those actions, the agent should obtain the user's permission, especially for the actions that are ``expensive'' in some sense (expend resources, are irreversible, might violate user preferences). When this is not the case, the agent should perhaps just go ahead, e.g., to minimize overload (X3).
Suppose the agent doesn't communicate this crucial information. In that case, it might hurt common ground between the user and the agent because they would not have had the opportunity to provide feedback to the agent. This feedback would allow the agent to achieve the goal more successfully. It might be beneficial even that user input is required before the agent executes any action.  

This challenge has become prominent with modern agents due to advances in generative and tool-using capabilities.  For example, when agents built on foundation models with general capabilities execute complex plans involving sophisticated tool use, it is unclear at what level of detail the agent should communicate. The optimal level of detail may depend on many factors, including the user's current understanding of the agent's behavior. For example, if there are steps that the user already expects, it may be redundant to convey them in too much detail. In contrast, steps that may be considered surprising may be more salient for communicating up front. This would allow the user to inspect them or change them if needed. This challenge may become even more important as agents become able to complete actions in real time because of advances in software and hardware for LLM inference, possibly making it harder for users to keep up with the agent's plan.

This challenge is related to work in the automated planning literature around {explainable planning} where the goal is to explain, e.g., why the agent's plan contains a particular action, why it did not prefer a different plan, or why the agent's plan in optimal~\cite{Chakraborti2020emerging,Kambhampati2021SymbolsAA}.

Some possible solutions to this challenge include presenting plans to the user and getting explicit approval---an approach implemented in the GitHub CoPilot for Workspace\footnote{\url{https://github.blog/news-insights/product-news/github-copilot-workspace/}}---allowing the user to engage in a dialog to inspect the plan, and modeling uncertainty around the user's mental model (e.g., in whether the user would be surprised) and using that to determine when to reach out to the user in more detail~\cite{Sreedharan2021FoundationsOE}.

\begin{tcolorbox}[breakable,colback=gray!5!white, colframe=gray!75!black, title=A2. What is the agent about to do?]
    
    \paragraph{Example 1: Data Deletion or Archiving} 
    
    A user has set up an agent to manage storage space on their server by automatically archiving or deleting old data. The agent is programmed to identify files that haven’t been accessed in a while and are likely no longer needed. Without consulting the user, the agent deletes many files it deems obsolete. However, these files include critical project documents the user didn’t realize were marked for deletion. The user had intended to review these files before making any decisions but wasn’t alerted by the agent about its planned action.
    
    When the user discovers the deletion, they realize that recovering the files is a complicated process that may result in permanent data loss. This scenario emphasizes the importance of the agent asking for explicit user permission before taking irreversible actions like data deletion. It also highlights the need for clear communication, especially when the consequences could be significant. The failure to communicate the deletion plan leads to unintended outcomes, misalignment with the user’s expectations, and potential operational disruptions.

    \paragraph{Example 2: Automatic Content Posting}

A user relies on an agent to manage their social media accounts, including drafting and scheduling posts. The agent is programmed to inform the user of its activities and decisions, sending frequent updates about each draft, scheduling changes, and analytics trends. While this level of detail aims to ensure transparency, it results in the user receiving a flood of notifications and updates throughout the day.

One day, the agent notifies the user of a minor change to a scheduled post amid dozens of other updates. Overwhelmed by the constant communication, the user overlooks the specific notification and misses an important detail: the agent had rescheduled the post to coincide with a sensitive industry event, inadvertently making it seem tone-deaf and sparking backlash.

This scenario illustrates the risk of over-communication, where excessive updates dilute the user's attention and reduce the likelihood of catching critical information. Striking a balance between transparency and concise communication is crucial to ensure the agent's actions align with the user's goals without contributing to information overload (X3).
\end{tcolorbox}
\subsection*{A3: What is the agent currently doing?}

While an agent is acting in an environment, how can the user understand what it is currently doing, the impact of the agent's actions that are in progress, and whether they should intervene to shape or halt the activities?  The distinction between this challenge and A2, stems from the time of communication. With A2 the communication occurs before the agent has executed an action while A3 concerns communication about an action being executed. This difference in timing suggests different potential solutions to A2 and A3. 

Consider, for example, an agent that can browse the internet (e.g., by operating a web browser) and also send and receive emails on your behalf. Suppose the agent was asked to plan and book your next vacation that fits your budget. To accomplish this goal, let's say that the agent pursued a multi-step approach---browsing the web to look up potential destinations and hotels, emailing relevant hotels at these destinations to negotiate prices, and finally using the web browser to book the final plan. How can the user, at the moment, stay aware of the agent's progress and verify the appropriateness of its actions? This information would allow the user to intervene on the spot to change the agent's course of action as necessary.

This problem can become much more challenging if the agent pursues strategies involving tens or hundreds of actions or executes steps in fractions of a second or in parralel.

 ``Unsafe'' or ``inappropriate'' agent behavior could worsen this challenge. For example, what if it emailed your boss to ask for a raise because it has discovered through a few steps of reasoning that an increased budget for the trip coming via a raise would help it to plan a better trip? \looseness=-1 

 Suppose the user cannot understand what the agent is currently doing. In that case, they cannot actively understand the agent's process and correctness or intervene to provide helpful guidance.  While existing human-AI guidelines do talk about the need to clarify which actions the system is taking to improve understanding, they do not delve into how to handle the level of complexity of modern agents. But there are some domains where this issue arises, such as with high-speed algorithmic trading, where trading systems may execute a large number of orders, or with semi-autonomous vehicles \cite{Heiden2017PrimingDB}. Some solutions developed in these fields may be relevant to human-agent communication, such as real-time updates and summaries of agents' behavior.

\begin{tcolorbox}[breakable,colback=gray!5!white, colframe=gray!75!black, title=A3. What is the agent currently doing?]
    
    \paragraph{Example 1: Online Shopping} 
    A user tasks an agent with purchasing supplies for their business, providing a budget and a list of required items. The agent begins the process by generating a plan, perhaps to compare prices at multiple online stores.  Challenge A2 reflects the need for the user to be able to easily correct such a plan before the agent starts execution. But even if the plan appears ok, as the agent rapidly executes these actions, it may encounter out-of-stock items or find the cheapest store won't actually ship the item until next month. Challenge A3 reflects the need for the agent to alert the user to potential problems or for the user to be able to effortlessly monitor the purchases in real-time to ensure they align with  preferences.

This scenario highlights the importance of real-time updates from the agent, enabling the user to stay informed about ongoing actions. Without effective communication from the agent, the user may miss the opportunity to intervene and correct potential mistakes before purchases are finalized, leading to unexpected costs or unsuitable products. Transparency in the agent’s actions is crucial to ensure the user’s preferences and constraints are respected throughout the shopping process.

\paragraph{Example 2: Web Navigation}

A user tasks an agent with gathering data to convey its current actions and intentions to the user actively. Unable to navigate the website successfully, the user decides to take an alternative route to achieve the goal. The agent autonomously drafts and begins to file a Freedom of Information Act (FOIA) request to obtain the needed information. 

While this scenario demonstrates how agents can creatively circumvent challenges, it also raises ethical or procedural concerns that may require user intervention. Moreover, it is crucial for the agent to actively convey its current actions and intentions to the user. For example, informing the user before drafting a FOIA request allows the user to assess the appropriateness of the action and intervene if necessary. Without transparency, the agent’s autonomous decisions might lead to unintended consequences, such as violating privacy, wasting external resources, or damaging the user’s reputation. Clear communication about what the agent is currently doing ensures that the user remains in control.

\end{tcolorbox}
\subsection*{A4: Were there any side effects or changes to the environment?}

How can users monitor crucial changes that agents make to their environment, such as cloud accounts, disks, or operating systems? For example, suppose a Web agent is tasked with booking a flight. What if it also signs the user up for a new credit card because it helps get a better deal? In these situations, the agent must convey any information about side effects that may be important for the user to know, particularly if they are surprising or violate social norms.

Previous literature on non-LLM agents, planning, and robotics has explored situations where systems might pursue plans that create side effects. Examples include reward hacking or irreversible actions. These issues could arise from poorly designed reward functions, incomplete models, or insufficient constraints~\cite{Amodei2016ConcretePI}.

With modern agents that may use tools in the open world, potentially accessing user accounts or computers, this challenge has become even more critical to address \cite{zhang2024interactionimpacts}. The complexity and the number of steps in their plans further exacerbate this issue. The challenge is further compounded when actions are executed quickly.

One approach to this challenge could be to implement post hoc explanations that summarize any significant alterations made by the agent to the environment, particularly irreversible ones. These explanations could help users stay informed about critical changes, ensuring they are aware of any unexpected or socially non-normative actions the agent takes. In some cases, potential side effects will need to be communicated before actions are taken so that the user can step in and intervene if needed (A2).

\begin{tcolorbox}[breakable,colback=gray!5!white, colframe=gray!75!black, title=A4. Were there any side effects or changes to the environment?]
    
    \paragraph{Example 1: File Backup and Archiving} 
A user tasks an agent with managing disk space by backing up and archiving old files that haven’t been accessed in a while. The agent optimizes for storage efficiency but inadvertently causes system-wide slowdowns because the archiving process consumes excessive processing power and memory. As a result, other applications become sluggish or unresponsive during the operation.

This scenario underscores the importance of considering and communicating the potential side effects of the agent's actions. While the primary task of archiving files was achieved, the lack of safeguards against system disruptions highlights the need for balancing task goals with broader operational impacts.

\paragraph{Example 2: Automated Web Research}
A user tasks an agent with conducting web research to gather information for a project, such as compiling data on market trends or finding academic articles. The agent autonomously browses various websites, downloading documents, saving web pages, and aggregating information from different sources. While performing these tasks, the agent inadvertently visits sites that contain malware or phishing attempts, which it unknowingly downloads along with the desired information. As a result, the user’s computer becomes vulnerable to security breaches, or sensitive data might be compromised.

In this scenario, the user is unaware of the potential risks the agent encountered during its browsing activities. The agent’s actions lead to unintended consequences, such as introducing malware into the system, which could have serious implications for the user’s data security. This example highlights the importance of the agent communicating its web browsing activities and any potential risks or side effects to the user. The user must be informed of the sites visited and the files downloaded, allowing them to review and mitigate any security threats before they cause harm.
\end{tcolorbox}
\subsection*{A5: Was the goal achieved?}

When users specify a high-level goal to the agent and the agent uses and executes a complex plan to solve it, the system needs to convey to the user information that allows them to verify whether the goal was achieved~\cite{Fok2024InSO}, and potentially how. This communication is important for the user to maintain an understanding of whether the agent correctly achieved the goal or whether the user needs to correct any mistakes.

How an AI system reached its goal has been studied in the automated planning literature, where explanation methods are developed to help users understand why the planner used a particular approach (instead of alternatives) to reach the goal~\cite{fox-arxiv2017}. Similarly, in the machine learning literature, when learned models are used to make (or recommend) critical decisions, explanations may be used to help users understand if and how the classifier arrived at a particular suggestion.

Confirming that modern agents have completed their goals to a user's satisfaction can be more challenging due to the complexity of requests that agents may be tasked with and the complexity and only partial observability of the real world. For example, automated planning approaches may assume that it's possible to specify the goal state, but in the open world, that may not be easily possible. Consider an agent that uses a complex plan to conduct literature surveys and eventually produce a 100-page report using a RAG-based approach. Verifying that the report accomplishes a user's intent is likely to require considerable effort. It may not be easy to validate automatically, e.g., because it may contain hallucinated facts and references.
Tool use further complicates the plan and action space---if the system did not include papers by a particular author in the report, did it check all possible publication sources?

Some possible solution directions to this challenge could include providing overviews and details of what the agent did, akin to the explainable planning literature, and allowing the user to drill down into the agent's activity. Agents could also provide affordances to map their decisions and outputs to primary sources, for instance, by providing citations.

\begin{tcolorbox}[breakable,colback=gray!5!white, colframe=gray!75!black, title=A5. Was the goal achieved?]
    
    \paragraph{Example 1:  Literature Review Compilation} 
    A user tasks an agent with compiling a literature review on a specific topic, instructing it to gather information from various academic sources, synthesize the findings, and produce a cohesive document. The agent uses a combination of web scraping, database queries, and natural language processing to collect and analyze relevant papers, extracting key points, statistics, and references. It then writes a literature review draft, complete with citations and a bibliography.

Once the agent completes the task, the user must verify whether the review meets the specified goals. This includes checking that the agent accurately represented the information, cited all sources correctly, and included the most relevant and up-to-date research. The user might find that the agent missed critical papers, misinterpreted some findings, or included references that are not credible. Additionally, the user must ensure that the review’s structure, argumentation, and conclusions align with the intended research goals. This scenario highlights the challenge of verifying whether a complex goal (like writing a literature review) has been fully achieved, particularly when the task involves multiple sources, tools, and steps. This may be easier if the agent conveys the sources and tools it used and the steps it took, perhaps on demand. It also emphasizes the importance of user involvement in reviewing and refining the agent’s output to ensure accuracy and completeness.

    \paragraph{Example 2: Event Planning and Execution}
    A user tasks an agent with planning and executing a corporate event, such as a conference or company retreat. The agent is responsible for managing all aspects of the event, including selecting a venue, arranging catering, booking accommodations for attendees, scheduling speakers, and coordinating with vendors. The agent uses various tools and services to handle these tasks, making decisions based on the user’s initial guidelines, such as budget constraints, preferred locations, and the event’s overall theme.

Once the agent completes the planning and begins executing the event, the user needs to verify whether the agent has achieved the goal effectively. This includes reviewing the venue choice to ensure it meets the event’s needs, checking that the catering arrangements accommodate all dietary restrictions, confirming that accommodations are appropriately booked, and ensuring that all speakers are scheduled correctly. The user might discover the agent overlooked critical details, such as failing to book transportation for out-of-town guests or scheduling conflicting sessions. Additionally, the user needs to ensure that the event stays within budget and adheres to the company’s branding and messaging guidelines. This scenario underscores the complexity of verifying whether a multifaceted goal like event planning has been fully achieved, particularly when it involves numerous interdependent tasks and decisions. It also highlights the necessity of user involvement in reviewing and potentially correcting the agent’s work to ensure a successful event.
\end{tcolorbox}

\section{Conclusion}
\label{sec:conclusion}
Generative language models have enabled complex tool-using agents, creating a new breed of AI systems that can execute actions that affect the digital and the physical world---often in sophisticated and sometimes unpredictable ways. These changes have made it increasingly difficult to establish and maintain common ground, a critical goal for effective collaboration. The difficulties of managing the complexity and rapid execution of agent actions, as well as the stochastic nature of their behavior, are new, emerging directly from these technological advancements. The long-standing challenges users face when interacting with AI systems---such as conveying goals, understanding agent competencies, and managing expectations---are increasingly complex with human-agent communication. At the same time, the capabilities of agent-based systems bring heightened stakes of successful communication versus failure. The new complexities coupled with increasingly high stakes have profound implications for transparency and trust.

Given these challenges, we issue an urgent call to action for the AI research community: there is critical need for sustained effort to develop new design patterns, guidance, and principles that prioritize enhancing common ground between users and AI agents. As many of the examples illustrate, the importance of developing and maintaining shared understanding cannot be overstated, as it directly affects the effectiveness and safety of AI systems, particularly in environments where the cost of failure is high. Maintaining common ground requires explicit and concerted efforts to create methods and tools that are aimed at establishing and maintaining shared understanding, thereby ensuring agents operate in a way aligned with human needs and expectations. 

The community must prioritize these challenges by fostering cross-disciplinary research and practical innovation as we advance. Only through these efforts can we build AI systems that are not only powerful but also transparent, trustworthy, appropriately controllable, and capable of maintaining effective communication about their activities with their human counterparts.

\section*{Acknowledgments}
We would like to thank Srini Iyer for helpful feedback on limitations of LLMs, Andrey Kolobov for feedback on connections to the planning literature, and our colleagues from Microsoft Research, the NORA research team, and the UW HAI group for many valuable conversations. 

\bibliographystyle{plain}
\bibliography{main}

\end{document}